\documentclass[prc,aps,nofootinbib,superscriptaddress,twocolumn,showpacs]{revtex4-1}
\usepackage{graphicx}
\usepackage{amsmath}
\usepackage{amsfonts}
\usepackage{bm}
\usepackage{xcolor}
\usepackage{ulem}
\usepackage{appendix}
\usepackage{verbatim} %block comment
\usepackage{enumitem} %allows customisation of list indentation
\usepackage{dcolumn} % for alignment on decimal point

\newcommand{\order}[1]{{\cal O}(#1)}

\newcommand{\bea}{\begin{eqnarray}}
\newcommand{\eea}{\end{eqnarray}}
\newcommand{\beq}{\begin{equation}}
\newcommand{\eeq}{\end{equation}}
\newcommand{\bqa}{\begin{eqnarray}}
\newcommand{\eqa}{\end{eqnarray}}

\def\mqo2{{\!\!\!}}

\newcommand{\piHB}{\pi^{\text{HB}}}

\begin{document}

\title{What different variants of chiral EFT predict for the proton Compton differential cross section---and why}
\author{Vadim Lensky}\email{Vadim.Lenskiy@manchester.ac.uk}
\author{Judith McGovern}\email{Judith.McGovern@manchester.ac.uk}
\affiliation{Theoretical Physics Group, School of Physics and Astronomy, University of Manchester, Manchester, M13 9PL, United Kingdom}
\author{Daniel R.~Phillips}\email{phillips@phy.ohiou.edu}
\affiliation{Institute of Nuclear and Particle Physics and Department of Physics and Astronomy, Ohio University, Athens, OH\, 45701, USA\\}
\author{Vladimir Pascalutsa}\email{vladipas@kph.uni-mainz.de}
\affiliation{ Institut f\"ur Kernphysik, Johannes Gutenberg-Universit\"at Mainz, J.-J.-Becher-Weg 45, 55128 Mainz, Germany}

\date{\today}

\begin{abstract}
We compare the predictions of different variants of chiral effective field theory for the $\gamma$p elastic scattering differential cross section. We pay particular attention to the role of pion loops, and the impact that a heavy-baryon expansion has on the behavior of those loops. We also correct erroneous results for these loops that were published in Ref.~\cite{PP03} (Phys.\ Rev.\ C {\bf 67}, 055202  (2003)).
\end{abstract}
\maketitle

Experiments to measure Compton scattering from the proton are presently being pursued at a number of facilities around the world, including MAMI (Mainz)  and HI$\gamma$S at TUNL.  Chiral effective field theory ($\chi$EFT) is one of the main theoretical techniques used to analyze $\gamma$p scattering data in the energy range $\omega_{\rm lab} \lesssim 350$~MeV. $\chi$EFT generates the most general Compton amplitude that is consistent with electromagnetic gauge invariance, the pattern of chiral-symmetry breaking in QCD, and Lorentz covariance, to any given order of the small parameter $P \equiv \{\omega,m_\pi\}/\Lambda$, with $\omega$ the photon energy, $m_\pi$ the pion mass and $\Lambda$ the breakdown scale of the theory. 

By now, $\chi$EFT calculations of the $\gamma$p amplitude exist in several different variants of the theory. Two key choices must be made: whether to include the $\Delta(1232)$ resonance explicitly,  and whether to maintain exact Lorentz covariance or not. The pioneering calculations of Compton scattering in $\chi$EFT~\cite{Be91,Be95} were performed in a theory with only nucleons and pions as explicit degrees of freedom: the effects of the $\Delta$ were encoded in a string of contact operators. This reduces the breakdown scale $\Lambda$ from its nominal value around the mass of the $\rho$ meson to the energy at which the $\Delta$ is excited, i.e. $M_\Delta - M_N \approx 300$ MeV. In addition this work employed the heavy-baryon (HB) expansion for the nucleon propagators, which amounts to making an expansion in $1/M_N$, alongside the EFT expansion in $P$.  In these calculations the polarizabilities $\alpha_{E1}^{\rm (p)}$ and $\beta_{M1}^{\rm (p)}$ are predicted at order $P^3$, and these predictions are in remarkably good agreement with present extractions from data~\cite{Dr03,Gr12}. The corresponding results for $\gamma$p observables agree moderately well with data for $\omega, \sqrt{|t|} \lesssim 150$ MeV~\cite{BGM,Be03,Be05}.  Beyond this domain the absence of the $\Delta$ significantly affects the ability of this $\chi$EFT variant to describe the physics, at least at this low order in the expansion.

Subsequent work~\cite{Bu92,He96,He97,He98,Hi03,PP03} showed how to incorporate the $\Delta$ as an explicit degree of freedom. In particular, Pascalutsa and Phillips~\cite{PP03} showed how to consistently resum the effects that generate the $\Delta$'s finite width within $\chi$EFT, and 
performed calculations of $\gamma$p scattering from threshold up to 350 MeV. 
It should be noted that adding the $\Delta$ as an explicit degree of freedom in $\chi$EFT means that the ratio $(M_\Delta - M_N)/\Lambda$ becomes one of the expansion parameters of the theory. Ref.~\cite{PP03} pointed out that this is numerically rather similar to the actual expansion parameter in $\Delta$-less calculations, $m_\pi/(M_\Delta-M_N)$, and denoted both as $\delta$; clearly $\delta\sim P^{1/2}$.   In $\delta$ counting powers of the electronic charge $e$ are shown explicitly, whereas the $\chi$EFT of Refs~\cite{Be91,Be95} counts $e\sim P$.  Thus the Thomson amplitude is $\order{P^2}\sim\order{e^2\delta^0}$ and structure effects start with $\pi$N loops at $\order{P^3}\sim\order{e^2\delta^2}$ in the low-energy region (the counting changes in the resonance region; for details see Ref.~\cite{PP03}).

More recently, Lensky and Pascalutsa~\cite{LP09} repeated and extended the calculations of Ref.~\cite{PP03} in a framework in which full Lorentz covariance in the Compton amplitude was maintained. They also 
incorporated additional effects such as $\pi \Delta$ loops, thereby extending the results to a higher order in the $\chi$EFT expansion.  Since $M_N > \Lambda$ the HB expansion should not harm the accuracy of the predictions, but  it has been pointed out that the differences between HB and covariant calculations can be marked---even at low energies---if both are only carried out to a low order in the expansion.  In particular, the computation of loops with the full nucleon Dirac propagator can soften the ultraviolet behavior of the integrand, leading to somewhat different predictions than those obtained in HB. Therefore in what follows we pay particular attention to how the HB expansion affects the predictions for the $\pi N$ loops for  $\omega_{\rm lab} \lesssim 350$~MeV.

The imminence of the aforementioned experimental data makes it timely to examine this issue, as well as other differences between these variants of $\chi$EFT, all of which are based on the same low-energy symmetries of QCD. Thus, in this brief report we collect the predictions of these different variants (with and without an explicit $\Delta$, with and without the $1/M_N$ expansion) for proton Compton cross sections. Readers who wish to learn more about $\chi$EFT in general, or the specifics of the different $\gamma$p calculations we are discussing, should consult the recent review~\cite{Gr12}, and note that a number of misprints present in Ref.~\cite{PP03} are corrected there.  There is also a discussion of the fact that 
the amplitude used in Ref.~\cite{PP03} for $\pi N$ loops had an incorrect analytic continuation above the $\pi N$ threshold, so that the cross sections given there above that energy are in error.  

The calculations of the $\gamma$p differential cross section presented below have the following ingredients.  All contain the nucleon Born graph (calculated with Dirac nucleons) and the $t$-channel $\pi^0$ pole graph (again covariant in all approaches).  The 
$\Delta$-full variants contain the $\Delta$-pole graphs ($s$- and $u$-channel) calculated as described in Ref.~\cite{PP03,LP09,Gr12}---covariantly and with a finite width stemming from $\pi$N loops.  When Compton pion-loop graphs are also added, the amplitude includes effects which are of leading or next-to-leading order throughout the kinematic region $0 \leq \omega_{\rm lab} \lesssim 350$ MeV (apart from  the loop correction to the $\gamma$N$\Delta$ vertices), and all effects up to next-to-next-to-leading order---$\order{e^2\delta^3}$---in the low-energy region $\omega_{\rm lab} \sim m_\pi$.

Pion-loop graphs in the Compton amplitude may be calculated covariantly or in the heavy-baryon ($1/M_N$) expansion, and will include $\pi\Delta$ loops if the $\Delta$ is included.  Our aim is to highlight the effects of the heavy-baryon expansion on the loop pieces of the amplitudes.  The different approaches make differing predictions for the polarisabilities at this order in the $\chi$EFT expansion; in particular the $\Delta$ pole makes a large contribution to $\beta_{M1}^{\rm ( p )}$. 

At next order, $\order{e^2\delta^4}$, $\alpha_{E1}^{\rm ( p )} $ and $\beta_{M1}^{\rm( p )}$ have counterterm contributions. We include these in \emph{all} the calculations presented here. In each case they are adjusted to yield 
the best-fit values of Ref.~\cite{Gr12}: $\alpha_{E1}^{\rm ( p )}=10.5\times10^{-4}~{\rm fm}^3$, $\beta_{M1}^{\rm ( p )}=2.7\times10^{-4}~{\rm fm}^3$. The corresponding values of the counterterms when dimensional regularization and the $\overline{\rm MS}$ scheme are used
are displayed in Table~\ref{table-cts}.  (The scale $\mu$ is chosen as $M_N$, where necessary.) The reader is referred to Ref.~\cite{Gr12} for a fuller explanation of all these different parts of the calculation.

The calculations presented here are as follows. 
{\noindent
\begin{enumerate}[leftmargin=*,itemsep=4pt,parsep=0pt]
\item Tree: the results for Compton scattering with nucleon and pion Born graphs, plus polarisabilities; all other calculations build on this one.  
\item +$\Delta$: Tree graphs plus the effects of the (dressed) $\Delta$ $s$- and $u$-channel pole graphs.
\item +$\piHB$: Tree graphs plus $\pi$N loops: the $O(e^2\delta^2)$ calculation in heavy-baryon $\chi$EFT without an explicit $\Delta$ degree of freedom.  
\item +$\pi$: as (3), but with  relativistic nucleon propagators in $\pi$N and $\pi\Delta$ loops.
\item +$\piHB,\Delta$: the $O(e^2\delta^3)$ calculation in heavy-baryon $\chi$EFT with an explicit $\Delta$, including tree graphs, $\Delta$ poles and HB $\pi$N and $\pi\Delta$ loops.
\item +$\pi,\Delta$: as (5),  but with  relativistic nucleon propagators in $\pi$N and $\pi\Delta$ loops.
\end{enumerate}}
Clearly only the last two are realistic calculations which can be compared with data in the resonance region (and indeed some way below).  For completeness we note that once $\alpha_{E1}^{\rm ( p )}$ and $\beta_{M1}^{\rm ( p )}$ are re-adjusted our cross sections with and without $\pi \Delta$ loops contributions are so close as to be not worth displaying separately.

The HB calculations depend on the frame chosen to define the pion-loop amplitudes while the covariant ones do not. The Breit frame has been chosen here, but in order to show the uncertainty this introduces, a grey band denotes the difference between the Breit and c.m.\ frame choices. Two caveats should be noted.  First, this is only an indication of the size of the  omitted $1/M_N$ contributions; in particular the two frames coincide at $180^\circ$.  Second, the lack of a band on other curves should not be taken to mean that they are somehow more precise; all are low-order calculations for which higher-order contributions of at least this relative size are expected.

\begin{table}[hbt]
\begin{tabular}{|c|c||c|c|}
\hline
$\alpha_{\rm EM}$ & $1/(137.04)$&
$f_\pi$ & $92.21$ MeV \\
$m_\pi^{\pm}$ & $139.57$ MeV &
$m_\pi^0$ & 134.98 MeV\\
$M_N=M_p$ & $938.27$ MeV&
$g_A$ & 1.270\\
%$\kappa^{\rm (s)}$ & $-0.120$\\ 
%$\kappa^{\rm (v)}$ & $3.706$ \\
$\kappa^{\rm (p)}$ & $1.793$ &&\\
$M_\Delta$ & $1232$ MeV&
$h_A\equiv 2g_{\pi N \Delta}$& $2.85$\\
$g_M$ & $2.97$ & $g_E$ & $-1.0$\\
\hline
\end{tabular}
\caption{Common parameters used by all calculations \cite{PDG12}. The $\pi^0$ mass is only used for the computation of the $t$-channel pion-pole graph. The $\pi$N$\Delta$ coupling constant $h_A$ is fit to the experimental $\Delta$ width and the magnetic and electric $\gamma$N$\Delta$ coupling constants $g_M$ and $g_E$ are taken from the pion photoproduction study of Ref.~\cite{PV06}.
% where the $\pi NN$ coupling $g^2_{\pi NN}/(4 \pi)=13.64$ is also employed. 
For definitions of all these symbols see Ref.~\cite{Gr12}.}
\end{table}

\begin{table}[hbt]
\begin{tabular}{|c|d|d|}
\hline
& \multicolumn{1}{c}{$\delta \alpha_{E1}^{\rm ( p )}$ ($10^{-4}~{\rm fm}^3$)} & \multicolumn{1}{|c|}{$\delta \beta_{M1}^{\rm ( p )}$ ($10^{-4}~{\rm fm}^3$)} \\
\hline
Tree & 10.5 & 2.7\\
+$\Delta$ &10.6 & -4.4\\
+$\piHB$ & -2.1 & 1.4 \\
+$\pi$ &3.6 &4.5\\
+$\piHB,\Delta$ & -9.8 & -7.1 \\ 
+$\pi,\Delta$ &-0.8 & -1.2 \\
\hline
\end{tabular}
\caption{Values for the counterterm contributions to $\alpha_{E1}^{\rm ( p )}$ and $\beta_{M1}^{\rm ( p )}$ in the different variants of $\chi$EFT considered here.}
\label{table-cts}
\end{table}

\begin{figure*}
\includegraphics[width=\textwidth]{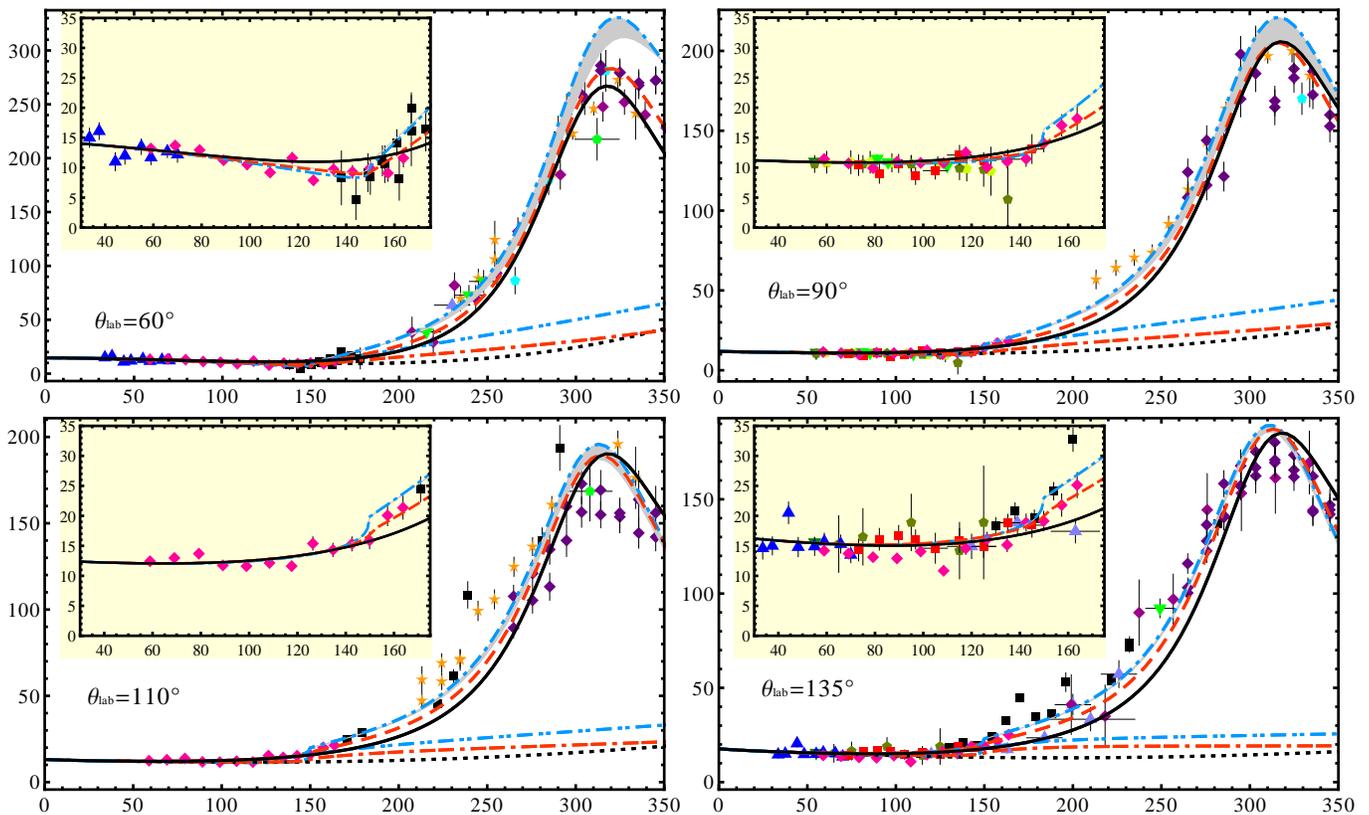}
\caption{Unpolarized differential cross section (in nb/sr) as a function of lab energy (in MeV) at fixed lab angles. Data from within $\pm 6^\circ$ of the quoted angle is shown in each panel.
 Our main results are the +$\piHB,\Delta$ (blue dash-dotted) and 
 +$\pi,\Delta$ (red dashed)  curves.  For comparison, we also show: +$\Delta$ (black solid),
+$\piHB$ (red dot-double dashed), +$\pi$ (blue dash-double dotted) and Tree (black dotted) calculation---for explanation of these terms and of the grey band, see the text.
The magenta/purple diamonds are data from Refs.~\cite{OdeL} and \cite{Wolf} as well as other experiments at MAMI, while the black squares are the data of Hallin et al.~\cite{Ha93}, and the yellow stars are from Ref.~\cite{Blanpied}. For the definition of other symbols see Ref.~\cite{Gr12}. }
\label{fig:xs_E}
\end{figure*}

The results of these calculations for the differential cross section at lab angles of $60^\circ$, $90^\circ$, $110^\circ$ and $135^\circ$ and for energies from threshold to the $\Delta$ peak are shown in Fig.~\ref{fig:xs_E}. The insets show, in detail, the behavior of the predictions around the $\pi$N threshold.  There is very little difference between the ``+$\pi, \Delta$" and ``+$\piHB, \Delta$" results at backward angles and higher energies. This is because the $\pi$N loops themselves are small in this regime. The $\pi$N loops are larger at forward angles for $\omega_{\rm lab}=300$--350 MeV, and the difference between the two calculations is more noticeable there: the ``+$\piHB, \Delta$" calculation overshoots the forward-angle data markedly. (We used the same $g_M$ in all calculations and this overshoot could partly be compensated by choosing a smaller value for $g_M$; in Ref.~\cite{Gr12} $g_M=2.85$ was found to give the best fit to Compton data.)  These forward- and backward-angle trends continue if one examines cross sections at angles less than $60^\circ$ or greater than $135^\circ$. 

One feature that is rather independent of angle is that the cusp that results from the opening of the $\pi$N channel is significantly stronger in the HB calculation. Measured relative to the ``+ $\Delta$" calculation, the cusp from the HB $\pi$N loops at this order is approximately twice as strong as in the covariant calculation. The insets in Fig.~\ref{fig:xs_E} show that the present experimental database cannot discriminate between the two calculations. The difference  in the vicinity of the $\pi$N threshold continues into the region $\omega_{\rm lab} \approx 200$ MeV, where a noticeable difference between the two predictions can be seen, essentially independent of angle. However, the data is sparser there.  The lesser prominence of pions in the covariant calculation reflects the fact that the relativistic baryon propagator gives the loops a softer high-energy behaviour than in the HB case.
(In actuality, the difference between the calculations with and without HB in the vicinity of the $\pi$N threshold is drastically reduced when the $\gamma$p $\pi$N loops are calculated to $\order{e^2\delta^4}$ in HB$\chi$EFT~\cite{McG01,McG12}.)

\begin{figure*}
\includegraphics[width=\textwidth]{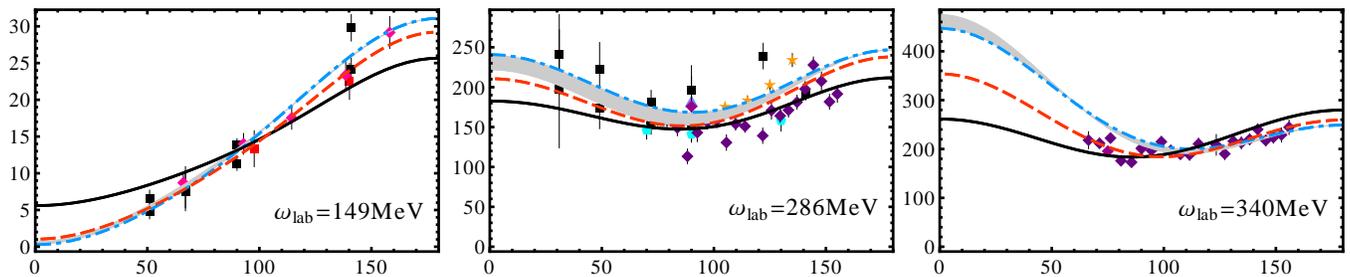}
\caption{Unpolarized differential cross section (in nb/sr) as a function of c.m.\ angle (in degrees) at fixed energies. Data from within $\pm 5$ MeV of the quoted energy are shown are each panel.
In this figure only three theory curves---``+$\Delta$",  ``+$\pi,\Delta$" and ``+$\piHB,\Delta$"---are shown. For legend see caption of Fig.~\ref{fig:xs_E}. }
\label{fig:xs_A}
\end{figure*}

In Fig.~\ref{fig:xs_A} we examine the cross section as a function of c.m.\ angle at three energies. The highest energy shown in Fig.~\ref{fig:xs_A} corresponds to $\sqrt{s}=M_\Delta$, and the trends already discussed in regard to the forward- and backward-angle cross sections are observed again there. (The rise of the ``+$\piHB,\Delta$" cross section at forward angles at this energy is tamed when the $\pi$N loops are computed to $\order{e^2\delta^4}$~\cite{Gr12}.) The situation is somewhat reversed at $\omega_{\rm lab}=149$~MeV (left  panel), where the forward-angle predictions of the HB and covariant calculations agree well with each other and with the trend of the data, but there is about a $10$\% difference at backward angles.

Overall then, the differential cross sections obtained in the two variants of $\chi$EFT that both include the leading $\pi$N loop effects and an explicit $\Delta$ are quite similar, provided the counterterms $\delta \alpha_{E1}^{\rm ( p )}$ and $\delta \beta_{M1}^{\rm ( p )}$ are included and adjusted to yield identical values for these scalar dipole polarizabilities. However, the values found for these counterterms in the $\chi$EFT variants considered here are rather different. They are particularly large in the ``+$\piHB\Delta$" calculation---especially as compared to the ``+$\pi,\Delta$" one---
which could be taken as evidence that the former $\chi$EFT expansion has less rapid convergence than does the latter~\cite{Hall:2012iw}.
 
Coming back to the cross sections, the differences between the
heavy-baryon and manifestly-covariant results are consistent with the fact that the latter calculation includes a number of mechanisms which would be higher-order in the HB calculation. Since $M_N \sim \Lambda$ the results obtained in these two variants of $\chi$EFT differ by amounts which are representative of the size of higher-order effects in the EFT expansion, 
provided $\alpha_{E1}^{\rm ( p )}$ and $\beta_{M1}^{\rm ( p )}$ are adjusted to the same value in both calculations. 
Once this is done $\chi$EFT makes predictions for the Compton cross section which are independent---up to  corrections of the expected, higher-order, size---of whether the heavy-baryon expansion is invoked or not. In light of this it will be interesting to examine the $\chi$EFT predictions for the asymmetries which are the focus of ongoing experiments at MAMI~\cite{Hornidge} and HIGS~\cite{Miskimen}.

\begin{acknowledgments} JMcG and DRP are grateful to Harald Grie\ss hammer for discussions regarding aspects of these calculations.
This work has been supported in part by UK
Science and Technology Facilities Council grants ST/F012047/1, ST/J000159/1
(JMcG,VL), the US Department of Energy under grant
DE-FG02-93ER-40756 (DRP), and the Deutsche Forschungsgemeinschaft through the Collaborative Research Centre SFB1044 (VP). DRP thanks the University of Manchester's Theoretical Physics Group  for 
hospitality during this project.  VL is on leave of absence from ITEP, Moscow.

\end{acknowledgments}

\end{document}